\documentclass[10pt]{wlscirep}
\usepackage[utf8]{inputenc}
\usepackage[T1]{fontenc}

\usepackage{amsmath}
\usepackage{graphicx}
\usepackage[colorinlistoftodos]{todonotes}
\usepackage{amssymb}
\usepackage{amsthm}
\usepackage{float}
\usepackage{subfig}
\usepackage{stfloats}
\usepackage{flushend,cuted}
\usepackage{indentfirst}
\usepackage{mdframed}
\usepackage{geometry}
\usepackage{graphics}
\usepackage{caption}
\usepackage{flushend}
\usepackage[sort, numbers]{natbib}

\title{Movies Network as the Indicator of Globalization}
\author[1]{Zilu Yu}
\author[2]{Chen Chen}
\author[3,4,*]{Dianbo Liu}
\affil[1]{University of Massachusetts Lowell, Boston, Massachusetts, 01854, United States}
\affil[2]{Jiangxi University of Finance and Economics, Nanchang, Jiangxi Province, 330013, China}
\affil[3]{Harvard Medical School, Boston, Massachusetts, 02215, United States}
\affil[4]{Computer Science and Artificial Intelligence Laboratory,MIT, Cambridge, Massachusetts, 02139, United States}
\affil[*]{Correspond to: dianbo@mit.edu}

\begin{abstract}
The main research involving globalization nowadays is to describe the impact of globalization in their respective fields. However, globalization is a complex phenomenon across multiple sections. But as a concept in the social science, it barely has the rigid mathematical foundation. Because of this lack, this article made a simple attempt to express and prove the trend of globalization with mathematical features. By abstracting an sub-area that is widely influenced by globalization, the article are trying to test whether this area can be used as an indicator of globalization.
\end{abstract}

\begin{document}
\twocolumn
\flushbottom
\maketitle
%
%
\setlength{\parindent}{2em}

\section*{Introduction}
\begin{figure*}[hb]
\centering
\begin{minipage}[c]{0.5\textwidth}
\centering
\includegraphics[height=4.5cm,width=6.5cm]{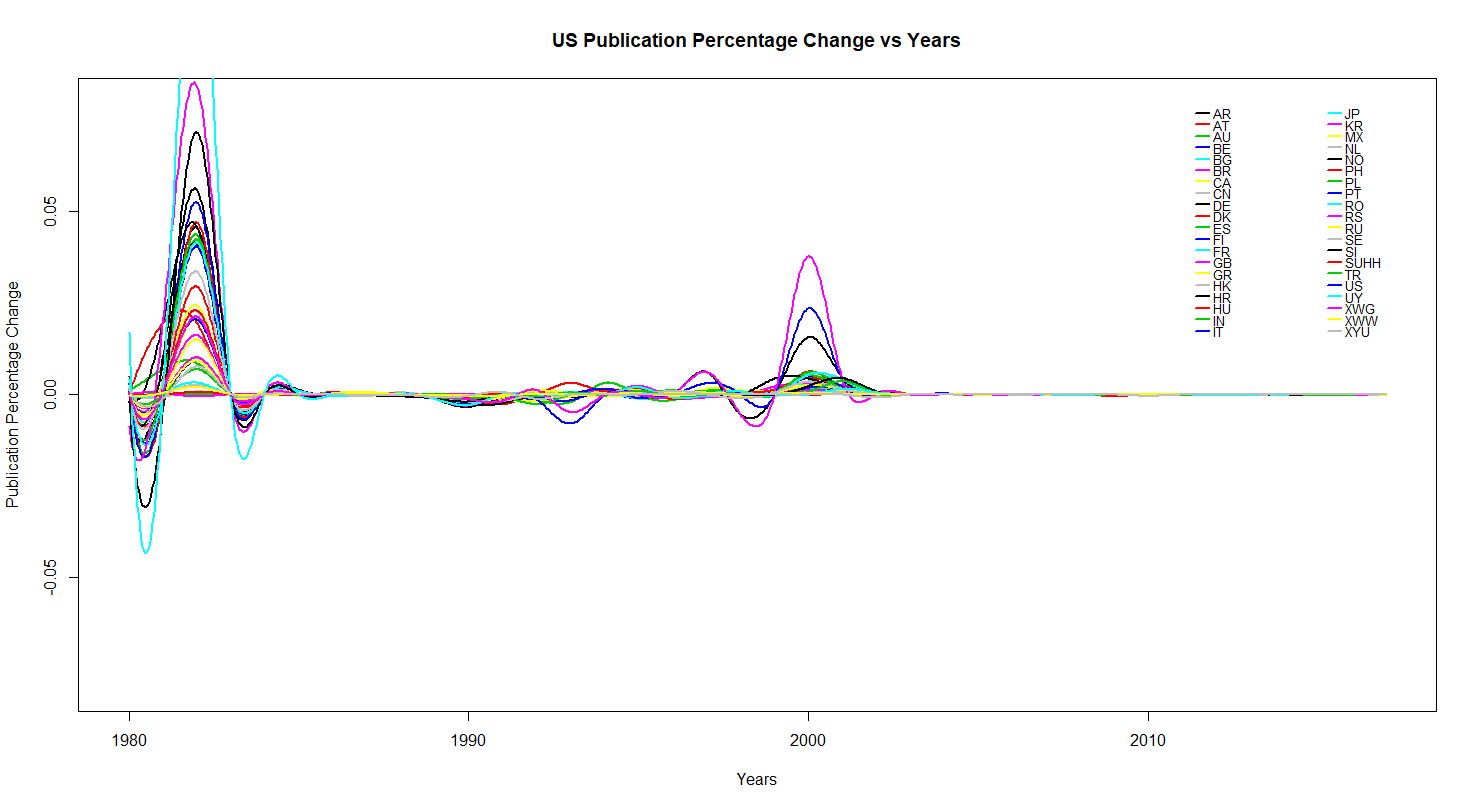}
\end{minipage}%
\begin{minipage}[c]{0.5\textwidth}
\centering
\includegraphics[height=4.5cm,width=6.5cm]{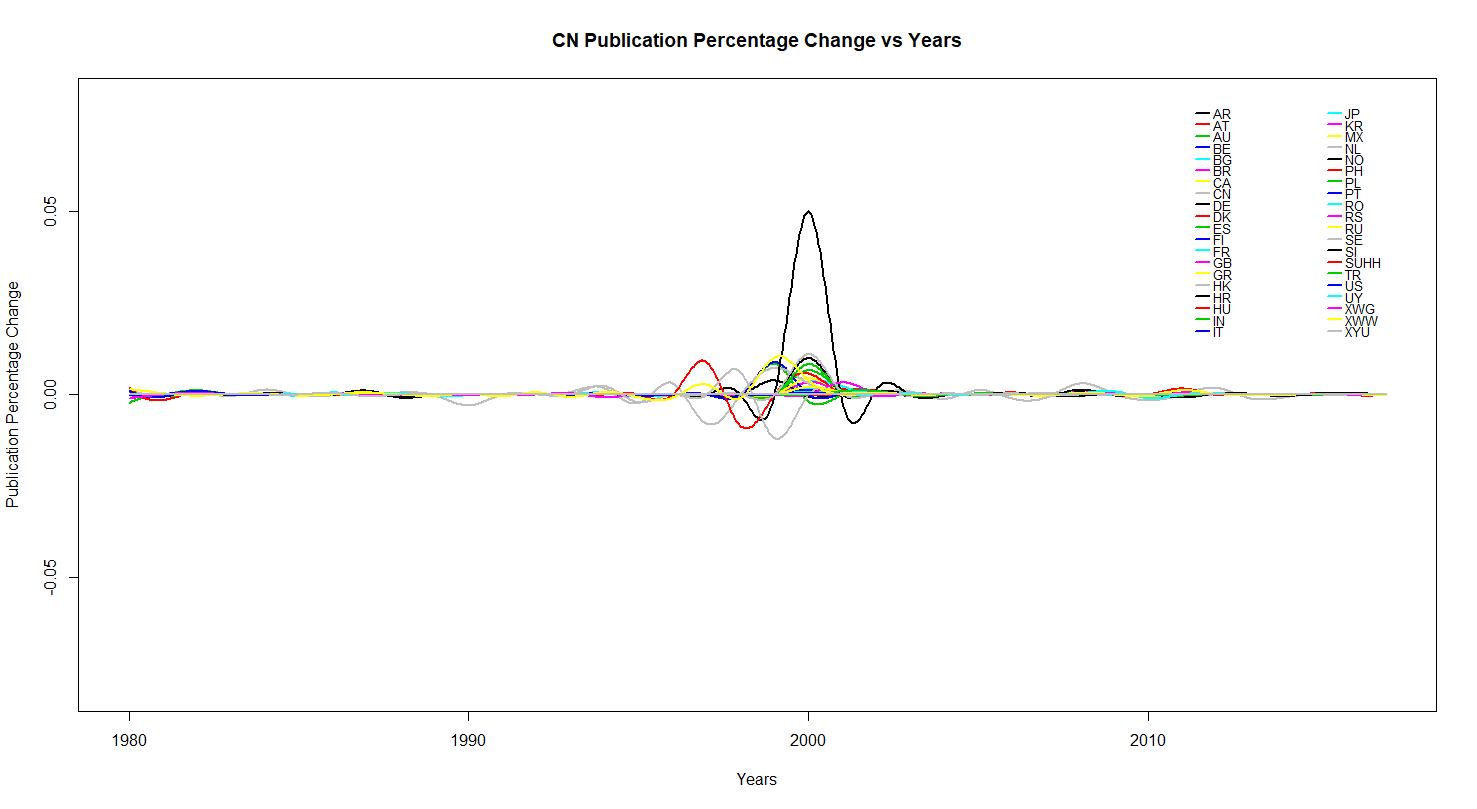}
\end{minipage}
\caption{\textbf{Percentage increase of the United States and China since 1980.} There are two peaks in the United States and a peak in China. The second peak of the United States and the peak of China appears simultaneously.}
\label{Figure1}
\end{figure*}
With the wave of globalization in the later period of the last century, human economic and cultural activities across large geographical areas have brought about an opportunity to study the trends of human behavior. Due to the expansion of distance, social structures, based on individuals, companies, and countries, are more likely to be abstracted\cite{kumartrawling,liu2017balance,liu2017integrative}; this also makes abstract analysis more significant.

In the field of economics, there is an indicator to describe economic changes. What these indicators have in common is to use a small range of numbers to represent a wide range of economic trends\cite{wankel2009encyclopedia}. However, unlike the economy, although globalization has been widely discussed and advanced over the past few decades, there are no effective indicator for globalization. So, if we imitate one of the specific indicator in economics, the coincident indicator, can we find an indicator that describes the overall globalization trend from a partial perspective? As a composite of culture and economy, in the general understanding, the legal publication of movies is a result of globalization, especially for American culture\cite{pells2004modernism}. Since United States is undoubtedly the core and leader of last century's globalization, American film publishing can be a potential object which can satisfies the requirement.

In order to abstract the film industry, different individuals, companies, or even countries, as a integration of personality\cite{ruesch2017communication}, can be considered as nodes in the network, so the structure of the film industry can be simplified into objects that can be analyzed. If it is assumed that the impact of each single link through the movie is the same, each movie will represent a edge between them, which make analyzing such social networks possible since scientific statistics and calculations can be applied on the network\cite{scott2017social}. Despite analyzing the factors that affect the network, we can find out the rules of the film industry or social development through appropriate inductions and make conjectures with verified hypothesis and use such generic rules\cite{smith1977database} to reflect the movement of globalization. Such networks with theoretical basis and realistic applications have fared extremely well in indicating the globalization trend  \cite{costa2011analyzing}.

Globalization has existed since ancient times, but this analysis is most interested in  the behavior of movie publication from the 1980s because the film industry was gradually developed from the beginning of the last century\cite{thompson2003film}. In the last century, globalization slowed down due to the Cold War, but it began to flourish in the 1980s\cite{ritzer2019globalization}. 

As another main character in globalization, China are also going to be one of the main objects in this analysis. Its network is set to check whether the property concluded from United States is able to apply to other cases. For example, it is expected that as China enrolled in globalization deeply, which shifted traditional interpersonal and international relationships in China\cite{yoshino2006social}, China also have the same behavior, but since China and United States enrolled into globalization at different times\cite{ritzer2019globalization}, if the film network can be used as an indicator of globalization, there should be a gap between their time.

In addition to descriptive words, since the property of networks is to be used as an indicator, mathematical proof is also needed. Since it is a comparison of a number to a series of numbers, it is reasonable to introduce the Wilcoxon test rather than typical paired t-test\cite{lowryconcepts}\cite{wilcoxon1945individual}.

\section*{Results}
When we briefly observed the trend of film publishing of United States in Figure \ref{Figure1}, it can be observed that the United States had a huge growth in 1982, and there was a small increase in 2000. Also, it is worth noting that in the remaining years, the percentage change is approaching zero. 
\begin{figure}[hb]
\centering
\includegraphics[height=4.5cm,width=6.5cm]{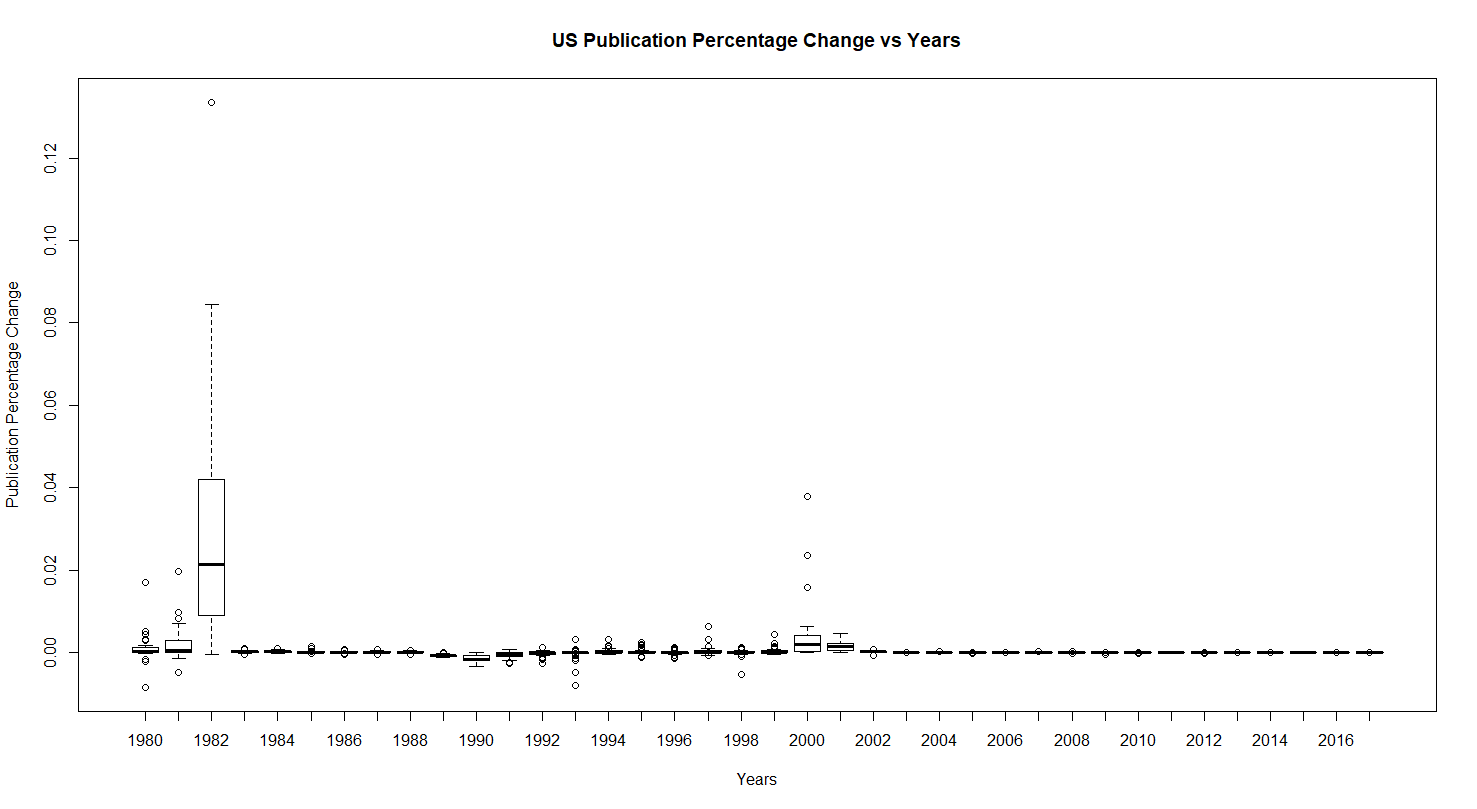}
\caption{\textbf{Box-plot of percentage change of the United States since 1980.} It can be seen that in 1982 and 2000, the distribution of data differed from the case where the other years were distributed around zero.}
\label{Figure2}
\end{figure}

In order to demonstrate the percentage change in the United States more intuitive, rather than using the ambiguous curve chart, box-plot, in Figure \ref{Figure2}, is applied to verify the fundamental observation described above. Therefore, after the United States 1982 film publications maintain a stable level in globalization since their film publication overlapped with other countries as the same degree. This fact may mean that as long as globalized, the relationship between the United States and rest of the world has not been reduced.  

\begin{figure}[htb]
\centering
\includegraphics[height=4.5cm,width=6.5cm]{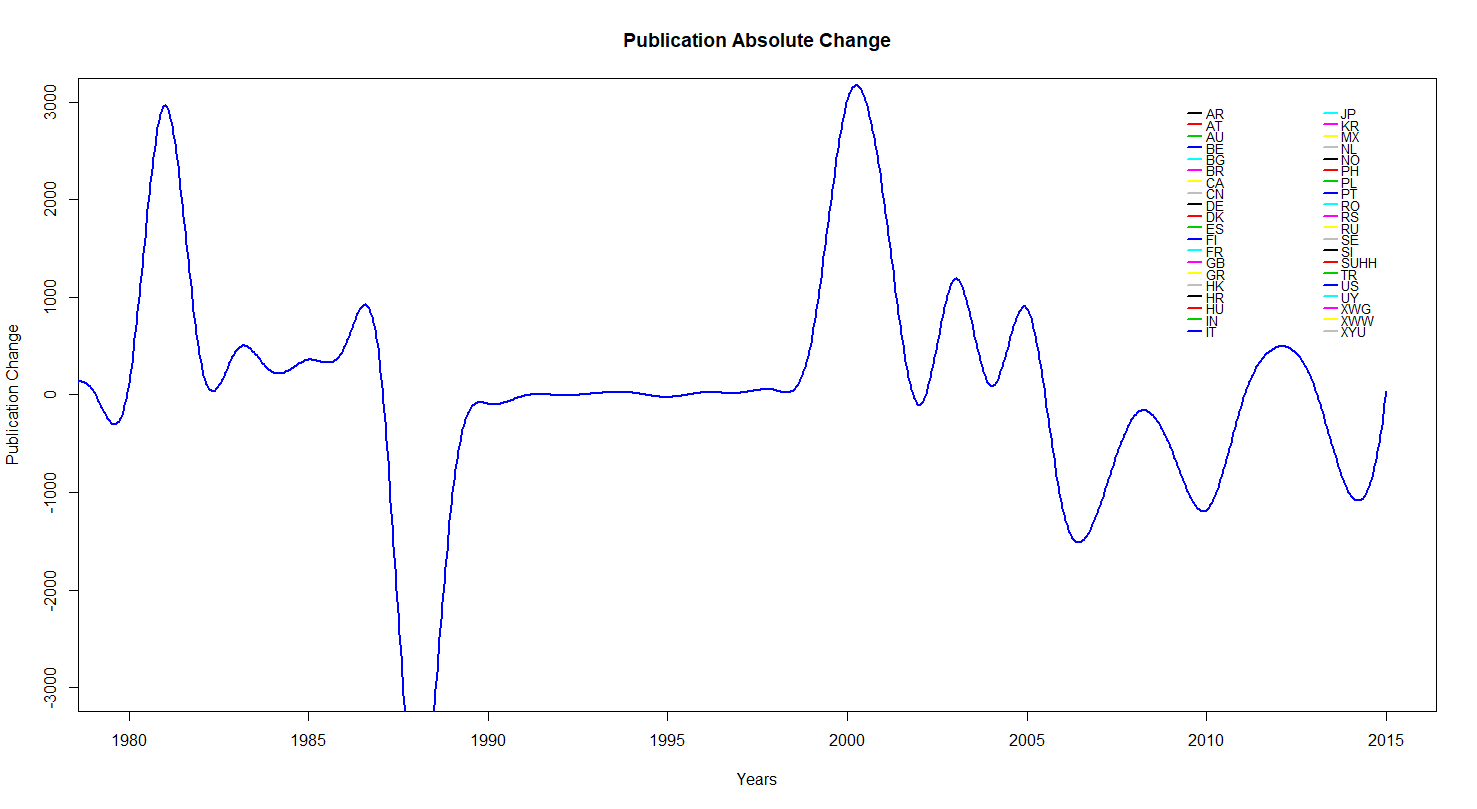}
\caption{\textbf{The absolute increase of the United States since 1980. }Can observe a huge decline around 1985 and a number of significant declines in the past fifteen years.}
\label{Figure3}
\end{figure}

For confirming the degree of globalization of the United Sates, the absolute change of the United states is checked. As shown in Figure \ref{Figure3}, obviously, the number of publications in the United States has decreased several times after 1982. In other words, the prosperity of the film industry does not actually affect the degree of globalization. So, it should be said that this analysis is a good way to avoid the impact of the film industry recession as considering its network as a globalization indicator.

After the preliminary thoughts are obtained through the observations and comparisons above, the mathematical proof needs to be considered. In this analysis, Wilcoxon test is applied to quantify the p-value of the data, which is average percentage change. As a result, data in year 1982 compared to that in rest years from 1980 to 2017, p-value is 0.0526 (<0.06); therefore, there exists a relevant significance. 

\begin{figure}[hb]
\centering
\includegraphics[height=4.5cm,width=6.5cm]{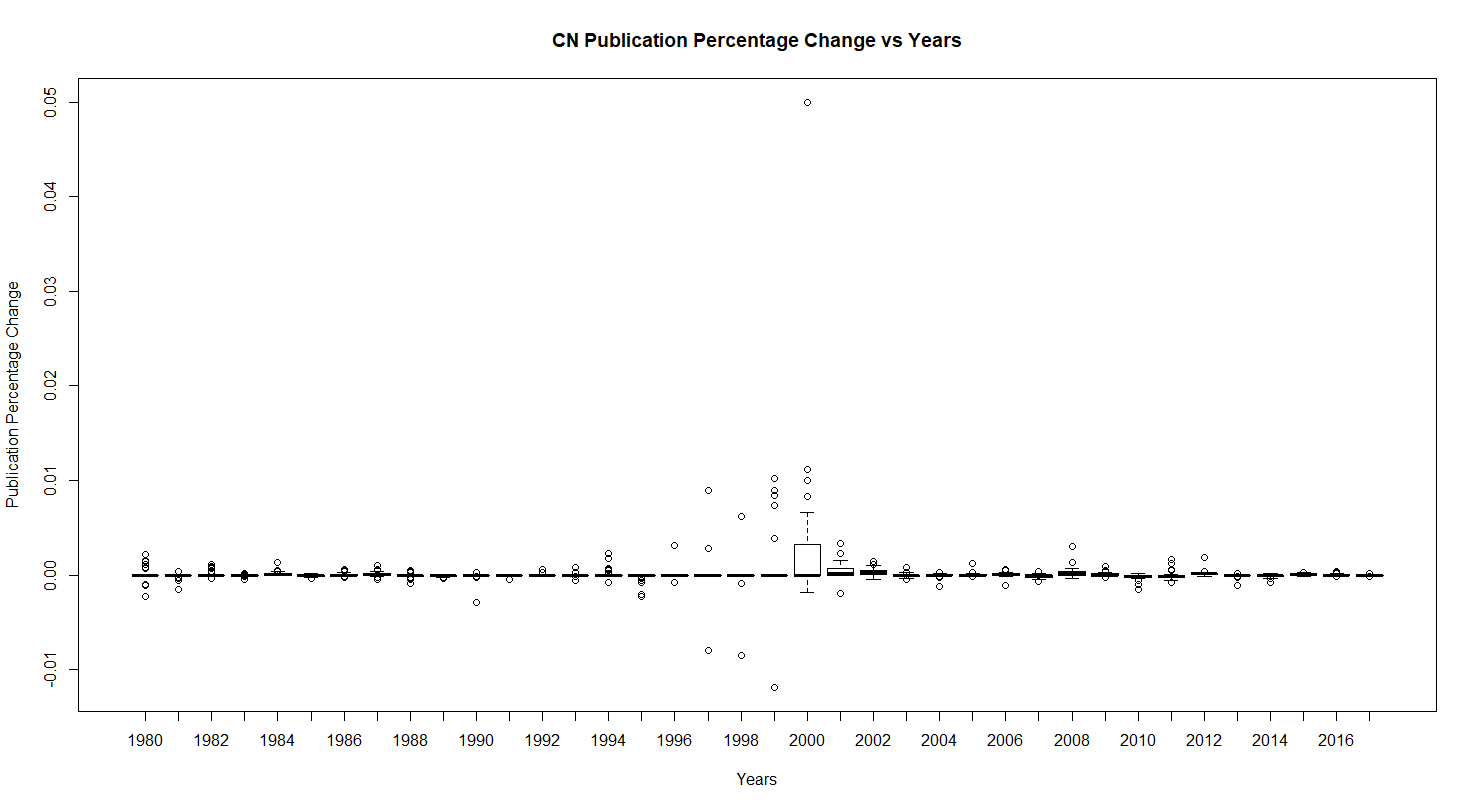}
\caption{\textbf{The absolute increase of China since 1980.}}
\label{Figure4}
\end{figure}

When the same research method is used in another country, China, that benefits in globalization\cite{guthrie2012china}\cite{milanovic2016global}, the results coincident with general understanding can be generated. Specifically, when applying the same percentage formula for China’s movies publishing network, there was, in Figure \ref{Figure1}, an obvious peak in 2000. Similarly, the box-plot in Figure \ref{Figure4} can demonstrate a more readable observation of its difference.
For the more rigorous mathematical meaning, the Wilcoxon test is also used on the average percentage change; in 2000, p-value of data compared to that in other years from 1980 to 2017 is 0.0526 (<0.06).

\section*{Discussion}
The brief method presented in this analysis is because, unlike the large number of mathematical applications in the economy\cite{ballard2004basic}, we have barely found articles that quantify the degree of globalization. Due to the lack of related research, the method used here still has many shortcomings since it is almost impossible to refer to previous studies.

The first limitation is that whether this method can be widely applied. This article only analyzes two typical countries since the United States and China are two countries that are vigorously enrolled in and benefited from globalization\cite{schaller2002united}, but it is unknown that whether is method is effective while applying it to other countries.

Despite the number of samples, the trend of these two countries is monotonous. If for a country that first involved in and then withdrew from globalization because of economic or political reasons, and this method clearly showing the growth and decline of degree of globalization, it will make the method more convincing. For example, Cuba, who used to be open but now resist to globalization\cite{baez2004state}, is probably a potential exemplar, but there is lack of its information.

In addition, in this article, the quantitative method we designed is based on a simple equation\eqref{1.1}. This is not a very clever equation, since it only avoid the error caused by the effect of different amounts of publication in the two countries.

Except the potential improvements in detail, there are still many room for improvement in how to describe globalization in mathematics. This article can be understood as detecting the change of \textsl{density} of globalization and how it differ from others, but, perhaps, there is a more complicated but more accurate way of calculating the degree of globalization.

Meanwhile, in the process of trying to find indicators of globalization, it should not be limited in movies network, which is, if using the similar definition in economy, more like a lagging indicator\cite{kirkpatrick2010technical}, since for studying some isolated countries such as North Korean, subordinate sections in economic or military areas are more likely to get sufficient data\cite{song1990rise}. Besides, multiple choices of indicators helps get more accurate results. 

Furthermore, research should try hard to move lagging indicators to leading indicators, like the work in economy\cite{ringold2007american}. Since the lagging indicator like movies network barely has meaning except summary, the leading indicator, absolutely, can be used on economic or politics decisions.

\section*{Materials and Methods}

In this analysis, we focus on the publication of movies around the world and the data we use is based on the international movies' data base, which contains about four hundred thousand movies and is collected by IMDB.com, by using r-studio 1.1.453.

After simply dividing the films based on years, the further matrix can be produced as taking years as a parameter. Then we need to use the movie ID to unify the data, because IMDB provides each movie with a different local name in different countries, and they are connected by a unique ID.Then, for each year, create a matrix $M_{CountriesMatrix}$ with row names and column names which is the countries that have shown any movie in that year. In this matrix, count each entry once for any combination of countries of each movie in that year; for example, if a movie is published in country A and B, the entry of AB and BA will add one. Finally, store all $M_{CountriesMatrix}$ into a list $L_{CountryMatrix}$, which contains all fundamental information of analysis. 

The diagonal in $M_{CountriesMatrix}$ represents the frequency of movies in relevant country. Recording the country names and corresponding frequency in a list $L_{CountriesTimes}$ each year; then, similarly, store them in a new list $L_{CountriesTimesEachYear}$. In the meantime, accumulate the frequency of movies of each country and record them in $T_{CountriesTimes}$. Choose the total frequency greater than a specific number, we can get a list of countries' names $L_{CountryList}$, which is the focus of this analysis since their frequency is large enough for having statistical meanings.

Since the frequency values of different countries vary greatly, a data indicating the degree of change is required. In order to avoid a situation where the denominator is zero and consider both countries in two sides of a link simultaneously, we introduced a formula to calculate the rate of change in the relationship between the two countries each year as following:
\[
\rho = \frac{\Delta X_{i} Y_{i}}{X_{i-1} Y_{i-1}}\label{1.1}\tag{1.1}
\]
Here $\Delta X_{i} Y_{i}$ means the number of publication increase between Country X and Y. $X_{i-1}, Y_{i-1}$ means number of movie publication in Country X and Y in previous year.

According to the above formula, for any country, we can deduce its $\rho$ with other countries in $L_{CountryList}$ each year. Then, choose a country that is more prominent in globalization, and based on the annual $\rho$  with other countries,  we can plot smooth curve graphs as Figure \ref{Figure1}. 

Also, as we acquire the peaks along the time, for example, year 1982 and 2000 for the United States. First of all, we have to make clear that these two years clearly behave differently, so here we can simply use box-plot to describe the performance of different years clearer. After we get the two possible significant year 1982 and 2000, we expect to make a rigid proof that these two years are significant. Therefore, we apply Wilcoxon test here since we want to compare one number to a series of numbers. In addition, we use average $\rho$ of all countries in $L_{CountryList}$ rather than any single $\rho$ for avoiding the effect of extreme value. After applying Wilcoxon test, we take year 1982 as the beginning of expansion of globalization of the United States.

Finally, we take China's movies network as an application of the method used above to test whether the conclusions directly analyzed from the characteristics of the film network fit with the common sense of reality.
\bibliography{bib}
\end{document}